# Direct observation of the influence of the *As-Fe-As* angle on the $T_C$ of superconducting $SmFeAsO_{1-x}F_x$

**Gastón Garbarino**
European Synchrotron Radiation Facility (ESRF), 6 Rue Jules Horowitz 38043 BP 220 Grenoble Cedex

**Ruben Weht**
Gerencia de Investigación y Aplicaciones, Comisión Nacional de Energía Atómica (CNEA), Avda. General Paz y Constituyentes, 1650 - San Martín, Argentina
Instituto Sabato, Universidad Nacional de San Martín-CNEA, 1650 - San Martín, Argentina

**Amadou Sow, André Sulpice, Pierre Toulemonde, Michelle Álvarez-Murga, Pierre Strobel**
Institut NEEL, CNRS & Université Joseph Fourier BP166, 25 Avenue des Martyrs, F-38042 Grenoble Cedex 9 France

**Pierre Bouvier**
Laboratoire des Matériaux et du Génie Physique, CNRS UMR 5628, Grenoble Institute of Technology, MINATEC, 3 parvis Louis Néel, 38016 Grenoble, France

**Mohamed Mezouar**
European Synchrotron Radiation Facility (ESRF),6 Rue Jules Horowitz 38043 BP 220 Grenoble Cedex

**Manuel Núñez-Regueiro**
Institut NEEL, CNRS & Université Joseph Fourier BP166, 25 Avenue des Martyrs, F-38042 Grenoble Cedex 9 France

The electrical resistivity, crystalline structure and electronic properties calculated from the experimentally measured atomic positions of the compound $SmFeAsO_{0.81}F_{0.19}$ have been studied up to pressures $\sim 20 GPa$. The correlation between the pressure dependence of the superconducting transition temperature ($T_C$) and crystallographic parameters on the same sample shows clearly that a regular $FeAs_4$ tetrahedron maximizes $T_C$, through optimization of carrier transfer to the $FeAs$ planes as indicated by the evolution of the electronic band structures.

email: gaston.garbarino@esrf.fr

# I. INTRODUCTION

The discovery of superconductivity[1,2,3,4] in layered compounds of iron in tetrahedral coordination (TIC) with superconducting critical temperatures ($T_C$) reaching up to 56K, has refreshed the interest in high temperature superconductors (HTSC). In particular, in search of a common pattern for HTSC's, the similarities and differences between cuprates and TIC's are been now deeply investigated. One example is the influence of the angle between the cations and anions in the active planes, $CuO_2$ and $FeAs$, that can have profound impact on their electronic properties[5,6,7,8]. For cuprates, the rule of thumb says that flatter $Cu$-$O$-$Cu$ buckling angles (~ 180°) give higher $T_C$'s, e.g. flat plane record bearing Hg cuprates[9], 214 materials[10], with notable exceptions as 123 compounds due to their unique structure[11]. For TIC's it has been reported[12,13,14,15] that regular tetrahedral $AsFeAs$ angles (109.47°) yield the highest $T_C$'s. Different theoretical models[7,8] have been proposed to explain this dependence, all based on the great sensitivity of electronic band structure with atomic positions. In this work, we report a simultaneous analysis of superconducting properties, crystal structure and electronic energy band structures on the same $SmFeAsO_{1-x}F_x$ ($x \sim 0.19$) compound as a function of pressure. The fact that we use pressure as a control parameter on the same sample is of extreme importance considering that it is a clean mechanism to modify the structure that avoid spurious impurity or other effects associated with the atomic replacement. Our results verifies the relation between the $T_C$ and the optimum $AsFeAs$ angle and gives an explanation that notably differs from previously proposed ones.

# II. EXPERIMENTAL TECHNIQUES

$SmFeAsO_{1-x}F_x$ samples were prepared by a high pressure – high temperature treatment using a "belt" type high pressure apparatus. $Sm$, $Fe$, $Fe_2O_3$, $As$ and $SmF_3$ powders were mixed together as precursors and pressed in the form of cylindrical pellets. For the synthesis, the pellet was introduced in a home made boron nitride crucible which was surrounded by a cylindrical graphite resistive heater, and the whole assembly was placed in the pyrophyllite gasket. The samples were treated at $6 GPa$, 1000-1100 °C during 4 hours then quenched to room temperature. X-ray diffraction pattern at room pressure and temperature shows a very good sample quality with less than 0.5% impurity phases. Superconductivity was checked by resistivity and a.c. susceptibility measurement showing an onset of superconductivity at $54 K$. The fluorine composition was obtained through the comparison of the cell volume at ambient conditions of our sample, $V_0 = 130.89(5) Å^3$, with the literature data of the parent compound[16] and different doping values[17]. Assuming a linear dependence of $V_0$ with fluorine doping up to ~ 20%, the effective composition was estimated to be $x \sim 0.19$.

The X-ray diffraction studies were performed on the $SmFeAsO_{0.81}F_{0.19}$ powder samples at the ID27 high-pressure beamline of the European Synchrotron Radiation Facility using a monochromatic beam ($\lambda = 0.3738 Å$) focused to $3 * 2 \mu m^2$. The diamond anvil cells with 600μm cullet diamonds with stainless steel gasket and two different pressure media were used. In the low pressure range ($P < 3GPa$) a 4:1 Methanol-Ethanol mixture assure a detailed analysis of the pressure dependence of the structural parameters, while using Neon the hydrostatic conditions are guaranteed up to the highest applied pressure $P < 20GPa$. Pressure was measured through the shift of the fluorescence line of the ruby. All the structural studies have been done at ambient temperature. The diffraction patterns were collected with a CCD camera and the intensity vs. $2\theta$ patterns were obtained using the fit2d software[18]. A complete Rietveld refinement was done with the GSAS-EXPGUI package[19].

The electrical resistance measurements were performed using a Keithley 2400 source meter and a Keithley 2182 nanovoltmeter. Pressure measurements, $0.2 - 7GPa$ (between $4.2K$ and

$300K$), were done in a tungsten carbide Bridgman anvil apparatus using a pyrophillite gasket and two steatite disks as the pressure medium[20].

## III. RESULTS AND DISCUSSION

Fig 1(a) shows the $3GPa$ room temperature diffraction pattern together with its Rietveld refinement, that was obtained using a tetragonal $P4/nmm$ unit cell and refining scaling factor, lattice parameters, profile shapes and the atomic positions of the $As$ and $Sm$, while the $O/F$ site was considered fully occupied. The pressure evolution of the diffraction patterns up to $20GPa$ does not show any evidence of phase transition. This is a common behavior in other structure studies under pressure at room temperature of the same 1111 family[21].

The pressure dependence of the volume, $V$, lattice parameters, $a$ and $c$, of the unit cell and the $a/c$ ratio are given in fig. 1 (b) and (c), respectively. A $3^{rd}$ order Birch-Murnaghan equation-of-state was used to determine the bulk modulus $K_0$ and its pressure dependence $K'_0 = \partial K_0 / \partial p$ [22]. The unit cell volume at ambient pressure, $V_0$, was fixed at the measured value $V_0 = 130.89(5) Å^3$, while $K_0 = 88.9(8) GPa$ and $K'_0 = 4.2(1)$ were obtained by the fit. Similar $K_0$ and $K'_0$ values have been reported in other compounds of the same 1111 family [21,23]. In the inset of Fig. 1(c) the pressure evolution of the $a/c$ ratio is presented, and clearly, a monotonous function (particularly a linear one) can be used to describe its pressure dependence in all the studied range.

The pressure was gradually increased in the low $P$ range ($P < 3GPa$) in order to have a very detailed analysis of the compression of the structure and correlate it with the superconducting and transport properties.

To highlight any anomaly in the volume compression, the data is plotted on Fig. 1(e) in terms of the Eulerian strain[24], $f = [(V_0/V)^{2/3} - 1]/2$, and normalized pressure, $F = p/[3f(1+2f)^{2.5}]$, calculated by fixing $V_0 = 130.89(5) Å^3$. A clear kink is indeed observed at $f = 0.0037$ ($p \cong 0.6 GPa$), that can be correlated with a change in the structure towards a regular $FeAs_4$ tetrahedron and maximal $T_C$, as will be discussed below.

The evolution of the structural parameters is presented on Fig 1(d), showing that the thickness of the *FeAs* layer, the distances between in-plane *Fe* ions and between *Fe* and *As* ions (named *width FeAs*, *FeFe* and *FeAs* in Fig. 1(d), respectively) have a very similar evolution under pressure with a variation of ~ 4% at $20 GPa$. The thickness of the *SmO* layer slightly changes all over the pressure range; whereas the distance between the *FeAs* and *SmO* layers (named *Free Space* in Fig.1(d)) shows a reduction of 15% at $20 GPa$. Similar structural trends with pressure have been reported in other oxypnictides compounds and are characteristic of a layered structure[23]. At this point it is important to mention the two different linear dependence in the *Free Space* behavior below and above $P \approx 0.6 GPa$, that is correlated with the kink at $f = 0.0037$ in the $F - f$ representation.

Fig. 2 shows the temperature dependence of the resistance for different applied pressures up to $7 GPa$. Under compression the absolute value of the resistance decreases. $T_C$, defined as the onset of the superconducting transition, is shown in the inset of Fig. 2 as function of pressure. It presents a non monotonous pressure behavior with a maximum at $0.6 GPa$, followed by a linear decrease for higher pressure.

The present combination of the structural parameters obtained from XRD studies and the transport properties measurements, gives very strong information about the important parameter that controls the electronic properties, and in particularly in this case the $T_C$. Up to now, several reports have pointed out the relevance of the *AsFeAs* angle and the *As* height respect to *Fe* layer on . For example, Zhao et al.[15] follow both variations and their correlation as the CeFeAsO system is doped. Also Kimber et al.[14] describe the variation of the angle with pressure, and compare with results on the pressrue variation of $T_c$ published elsewhere on a different batch of samples. In both cases, the fact that the measurements are not done on the same samples diminishes the reliability of these reports. We follow the variation of $T_c$ and of the structure in the *same sample* , leaving no margin for uncertainties due to sample preparation, impurities, etc. It is in this respect that our work is more trustworthy. This is done in Fig. 3, that shows the pressure dependence of $T_C$ (panel (a)), the *AsFeAs* angle (panel (b)) and the *As* height (panel (c), defined by the distance between *As* and *Fe* planes, i.e. $hAs = [z(As) - z(Fe)]*c = [z(As) - 0.5]*c$. The data unambiguously prove that $T_C$ reaches its maximum of $55.2K$ at the same pressure as the *As* height increases up to the optimal value of $1.39 Å$ and the *AsFeAs* angle moves to its regular tetrahedral value of $109.47°$ (dashed line in fig.3 (b)). This is the first experimental evidence of a clear correlation between the structural parameters and the superconducting transition on the same sample by means of a clean technique like pressure.

A special discussion merits the study of the relation between the *As* height ($hAs$) and the *AsFeAs* angle ($\alpha$). A simple analysis of the tetragonal structure allows to deduce the relation between those two structural parameters: $\alpha = 2\tan^{-1}\left(\frac{a/2}{hAs}\right) = 2\tan^{-1}\left[\frac{(a/c)}{2\,zAs - 1}\right]$. Considering that the function $g(x) = \tan^{-1}(x)$ and the pressure evolution of the $a/c$ ratio (inset of Fig.1 (b)) are monotonous, it can be inferred that a non monotonous behavior in $\alpha$ is related with the opposite kind of behavior in $zAs$ or $hAs$; i.e. $\alpha \propto 1/zAs$. In addition, in the case of a non distorted $FeAs_4$ tetrahedron, the relation is reduced to $hAs = a/\sqrt{8}$. As the *a* lattice parameter is $a \cong 3.95 Å$ the optimum of value for $hAs$ is $1.39 Å$ and $\alpha = 109.47°$. This close relation between these two parameters imply that the structural effects on $T_C$ can be explained either by the angle $\alpha$ or the $As$ height $hAs$.

The relation between a regular $FeAs_4$ tetrahedron and maximum $T_C$ has been discussed in various numerical studies[7,8]. A two orbital band approach yields a flat band, sensitive to changes in the tetrahedron angle, with a strong peak on the density of states at the Fermi level[8]. Such a Fermi surface feature can affect both nesting, or, in a strong coupling approach, superexchange. Including a third orbital allows symmetry breaking through tetrahedron deformation. Considering the strong hybridization of the five $3d$ orbitals, it has been pointed out[7] that a five orbital model is necessary to analyze the effect of the $hAs$ on the spin fluctuations mediated superconductivity. In this framework, the structural parameters modify the nesting scenario between the different bands and consequently the gap function and the $T_C$. However, all these considerations may not be applicable here, as they rely on stronger angle variations than those actually measured in our $SmFeAsO_{1-x}F_x$.

In order to obtain a detailed microscopic interpretation of the subtle structural effects on the electronic properties and the superconducting transition, *ab-initio* calculations using the actual, measured pressure dependence of the crystal structure and the atomic positions were performed.

All the electronic structure calculations have been carried out using the full-potential linearized-augmented plane wave (FP-LAPW) method as implemented in the Wien2k code[25] based on the Density Functional Theory. We considered the PBE[26] generalized gradient approximations to treat the exchange correlation potential. Sm *5f* states were treated as pseudocore states, restricting them to the muffin-tin spheres (for this reason we have considered quite large muffin-tin radius for Sm, of 2.5 au.). Doping was simulated assuming a virtual oxygen atom with a charge of $8+x$ ($x=0.19$ in our case).

Band structures at three representative pressures ($0.1$, $0.5$ and $3.6 GPa$) are presented in Fig. 4. These three pressures correspond to conditions below, at, and above (see Fig. 4a, 4b and 4c, respectively) the optimal condition at which $T_C$ is maximized. The main result that we observe is the appearance of an electron band crossing the Fermi level for the regular tetrahedron that induces the appearance of an electron pocket around the Z point on the Brillouin zone. A careful analysis of the orbitals that give raise to this band shows that it is essentially a 3 dimensional band that correlates the $4f$ $Sm$ with the $2d$ $Fe$.

The studied sample has a $F$ content of $\sim 0.19$, i.e. a doping larger that the optimal one[27], and the sample is thus on the overdoped region. In this case, pressure would induce a monotonous reduction of $T_C$. At the optimum pressure, the $FeAs_4$ tetrahedron is not distorted and the $Sm$ band dips below the Fermi level, pumping electrons and reducing the doping towards the optimal one. In this way the maximum in $T_C$ can be simply explained in this case by charge transfer towards the optimal doping. A representative of the evolution of charges in the $FeAs$ layer is shown on Fig. 4d (red dots), where it is clear that there is a reduction of the "number of electrons" at the optimum pressure in the $FeAs$ block for the regular tetrahedron due to the lowering of the $Sm$ band. The "electron concentration" was also calculated for a structure with the same lattice parameters but an irregular tetrahedron (with similar angles to the low pressure case) that shows that in such a case the evolution is monotonous (blue diamond in Fig. 4d). Superconducting $T_C$ changes under pressure due to charge transfer under pressure have also been observed in 122 materials[28].

As our sample is in the overdoped region, there is no distortion to be expected below $T_c$, as reported for underdoped 122 materials[29]. Thus, our calculation using the room temperature atomic positions at each pressure, should be valid down to the lowest temperatures.

The decreasing measured pressure dependence of $T_c$ can be attributed to the combined effect monotonous charge transfer and electronic susceptibility damping. The contribution of the $Sm$ band just adds a small contribution that compensates this effect only around the ideal tetrahedron configuration. This small contribution on a monotonous dependence explains the maximum.

## IV. CONCLUCION

In conclusion, correlated structural, transport and electronic band properties of $SmFeAsO_{0.81}F_{0.19}$ have been performed. No phase transition was observed up to $20 GPa$ and the measured compressibility is similar to other compounds of the same family. This is the first pressure experiment done on the same sample where we show the importance of the structural parameters $As$ height and $AsFeAs$ angle on $T_C$. A maximum $T_C$ of $55.2K$ is attained at the same pressure where the $As$ atoms height is $1.39Å$ and the $AsFeAs$ angle reaches its non distorted value of $109.47°$.

It is furthermore shown that this $T_C$ maximum is due to an electron transfer to the $SmO_{1-x}F_x$ plane changing the electron concentration in the $FeAs$ planes from overdoped to optimally doped. As this picture is different from previously proposed scenarios, it suggests that the origin of the tetrahedron-$T_C$ relation can either have multiple origins or that the basic idea for the problem is still to be found. We hope that this work will stimulate theoretical groups to search for a general explanation, probably relying on symmetry, not details of bands calculations, that we suggest may be different for each compound.

## ACKNOWLEDGMENTS

The authors express their thanks to J. Jacobs for the preparation of the diamond anvil cells and W. A. Crichton for the gas loading. This work was partially supported by the project SupraTetraFer ANR-09-BLAN-0211 of the Agence Nationale de la Recherche of France. RW is member of CONICET-Argentina and gratefully acknowledges partial support from CONICET (grant PIP 112-200801-00047) and ANPCyT (grant PICT 837/07).

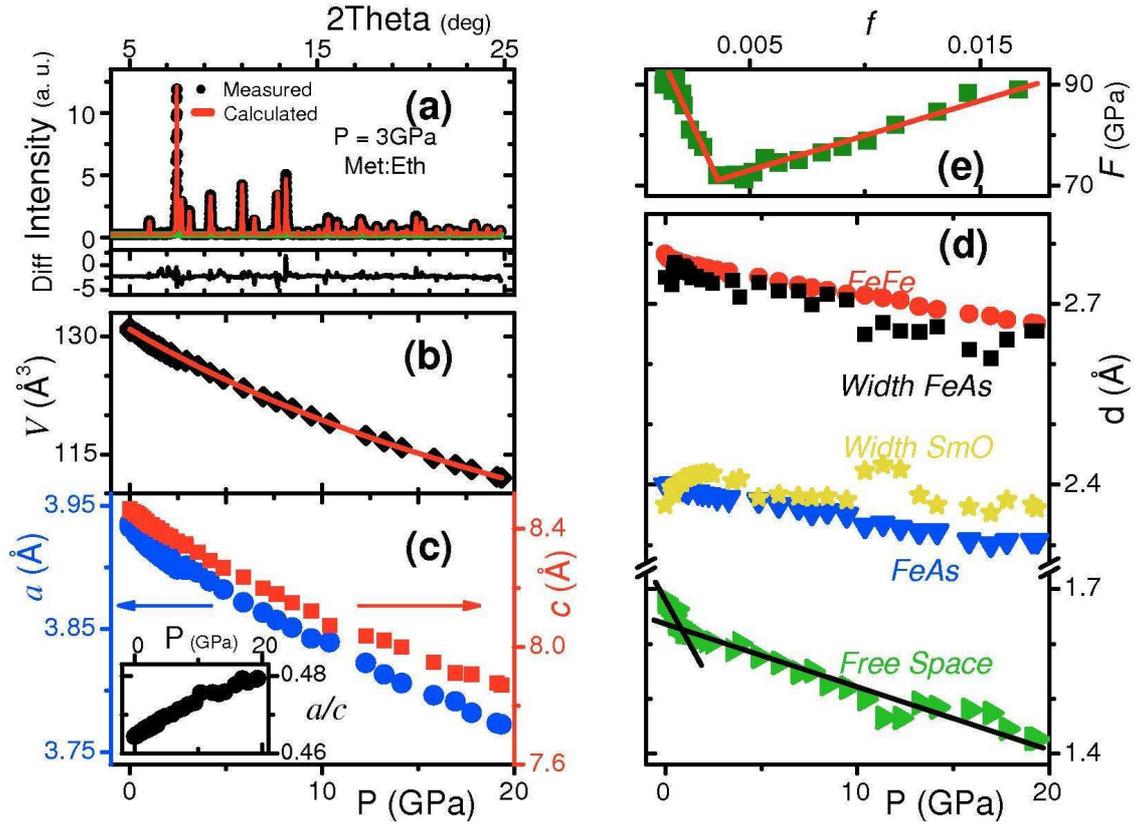

**FIG. 1. (Color online) (a)** Xray diffraction pattern measured at $3GPa$ of the $SmFeAsO_{0.81}F_{0.19}$ (black dots). The red line represents the Rietveld refinement using a tetragonal $P4/nmm$ structure, while the black line is the difference between the model and the data. Pressure dependence of the **(b)** volume, **(c)** lattice parameters and **(d)** interatomic distances of the $SmFeAsO_{0.81}F_{0.19}$ sample. The inset of panel **(c)** shows the monotonic evolution of the $a/c$ ratio. **(e)** Normalized pressure, $F$, versus eulerian strain $f$. The lines show linear fits of the regions of the compression curves.

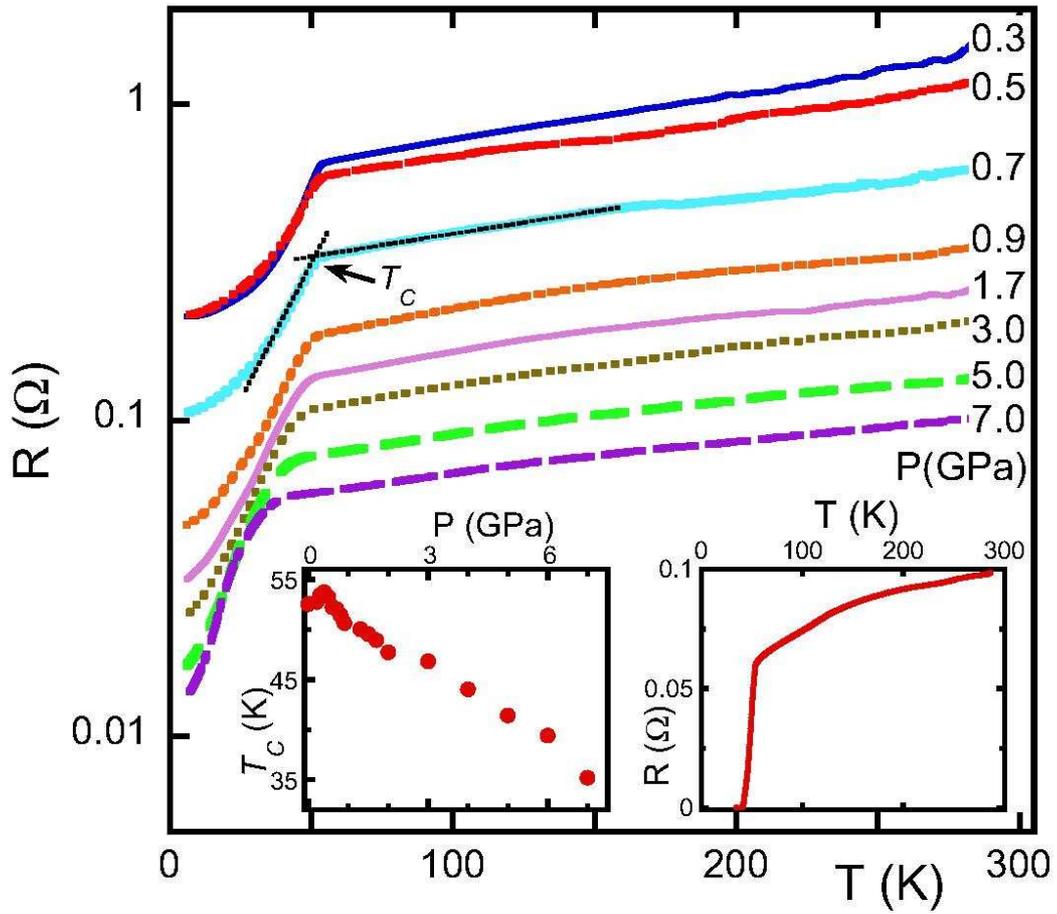

**FIG. 2. (Color online)** Temperature dependence of the resistance for different applied pressure of the sample $SmFeAsO_{0.81}F_{0.19}$. Left inset: the pressure evolution of $T_C$, defined as the onset of the superconducting transition (see intersection of the doted lines). Right inset: Resistance of the sample at ambient pressure.

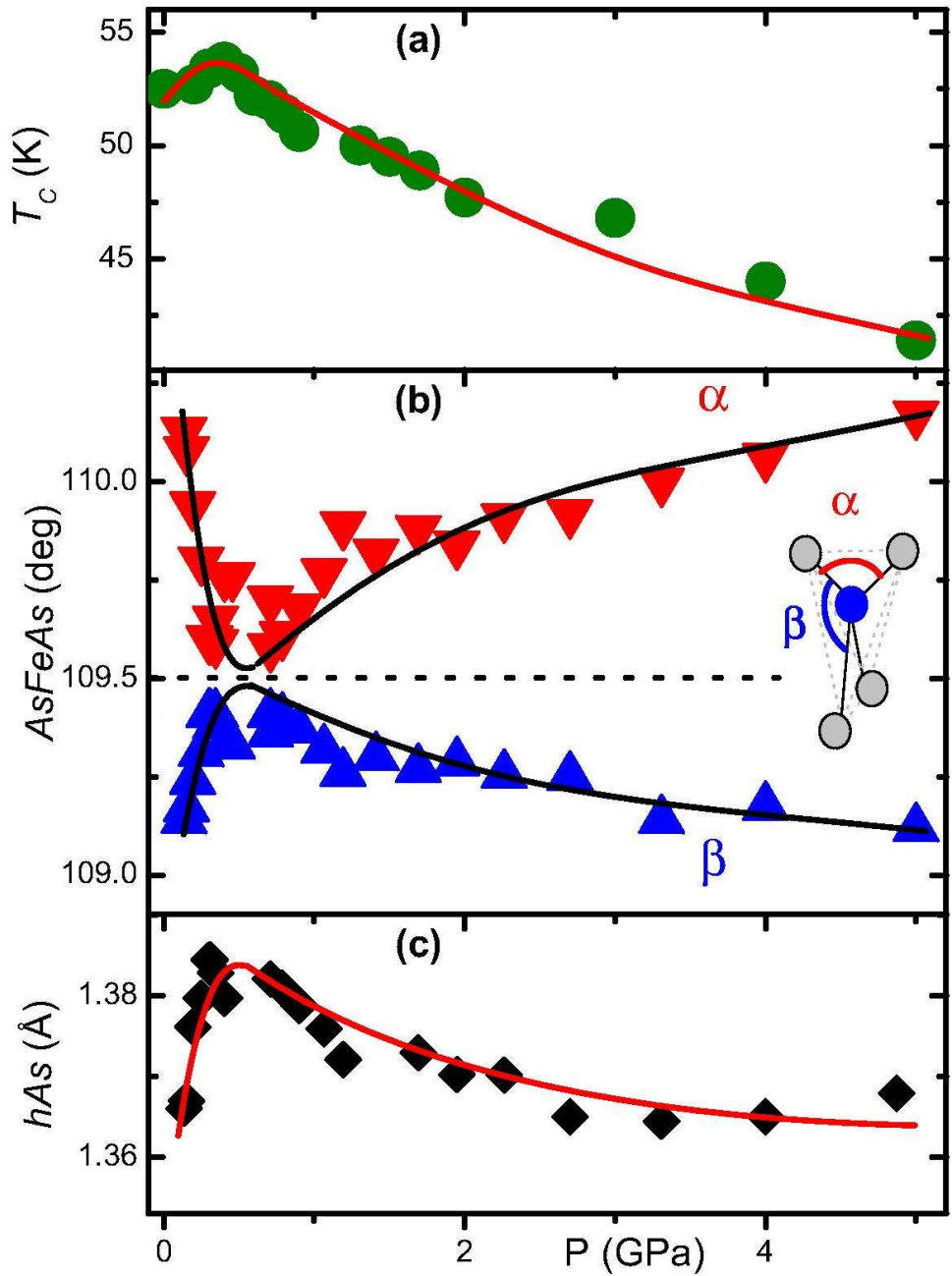

**FIG. 3. (Color online) (a)** Correlated evolution of $T_C$, **(b)** the tetrahedral *AsFeAs* angle and **(c)** the height of the *As* (*hAs*) atom relative to the Fe layer of $SmFeAsO_{0.81}F_{0.19}$. The maximum of $T_C$ clearly corresponds with the maximum *As* height and the regular tetrahedral angle ($\alpha = \beta = 109.47°$, represented with the dashed line in (b)).

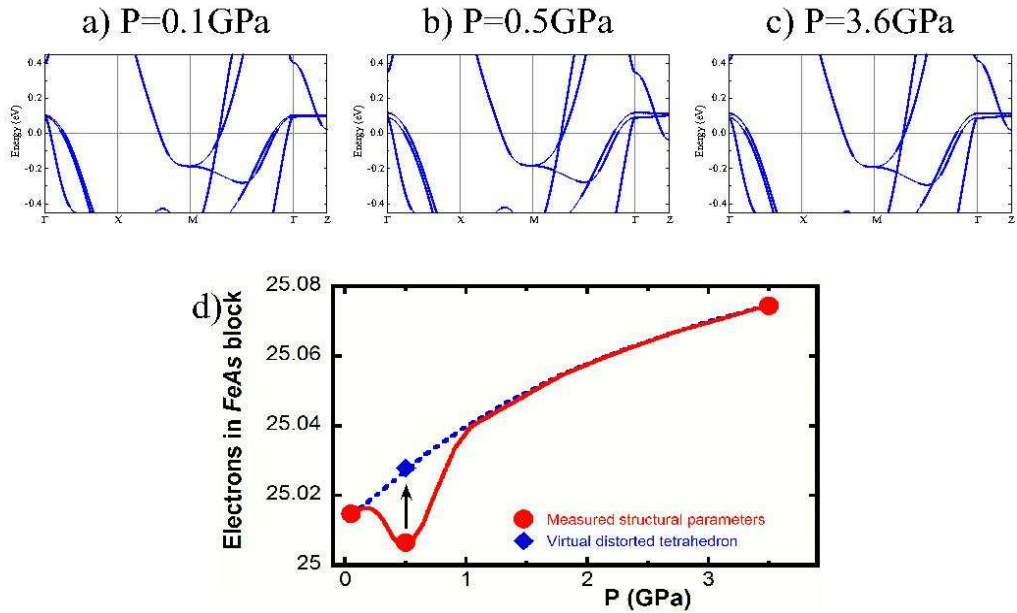

**FIG. 4.** (**Color online**) (**a**) , (**b**) and (**c**) Pressure evolution of the band structure corresponding to $0.1 GPa$, $0.5 GPa$ and $3.6 GPa$, respectively. (**d**) Evolution with pressure of the charge in the *FeAs* block (red dots) calculated using the measured structural parameters and a fictitious distorted tetrahedron (blue diamond). To evaluate this magnitude we use the corresponding LAPW projections inside the muffin-tin spheres. Lines are guide for the eyes.